\documentclass[twocolumn]{aastex6}
\usepackage{natbib}
\usepackage{amsmath}
\usepackage{index}
\usepackage{color}

\usepackage{rotating}
\usepackage{enumitem}
\usepackage{aas_macros}

\font\manual=manfnt at 7pt \def\dbend{\hbox{\raise0.9ex\hbox{\manual\char127\hspace{0.6em}}}}

\newcommand\Ion[2]{\ensuremath{\mathrm{#1\,\scriptstyle #2}}}

\newcounter{INTERNALionstage}
\providecommand{\ion}[2]{
  \setcounter{INTERNALionstage}{#2}%
  \Ion{#1}{\Roman{INTERNALionstage}}}

\def\gtsim{\mathrel{\hbox{\rlap{\hbox{\lower4pt\hbox{$\sim$}}}\hbox{$>$}}}}
\def\lesssim{\mathrel{\hbox{\rlap{\hbox{\lower4pt\hbox{$\sim$}}}\hbox{$<$}}}}

\DeclareMathAlphabet{\vib}{OML}{cmm}{m}{it}

\received{January 17, 2017}
\accepted{April 12, 2017}
\slugcomment{The Astrophysical Journal Supplement Series, 230:8 (12pp), 2017 May}

\begin{document}

\title{Atomic Data Revisions for Transitions Relevant to Observations of Interstellar, 
       Circumgalactic, and Intergalactic Matter}
\author{Frances H. Cashman\altaffilmark{1}, Varsha P. Kulkarni\altaffilmark{1}, 
        Romas Kisielius\altaffilmark{2}, Gary J. Ferland \altaffilmark{3},
        Pavel Bogdanovich\altaffilmark{2}}

\altaffiltext{1}{Department of Physics and Astronomy, University of South 
Carolina, Columbia, SC 29208, USA}
\altaffiltext{2}{Institute of Theoretical Physics and Astronomy, Vilnius University, 
                 Saul{\.e}tekio al. 3, LT-10222 Vilnius, Lithuania}
\altaffiltext{3}{Department of Physics and Astronomy, University of Kentucky, 
                 Lexington, KY 40506, USA}

\begin{abstract}
  Measurements of element abundances in galaxies from astrophysical  spectroscopy 
depend sensitively on the atomic data used. With the goal of making the latest 
atomic data accessible to the community, we present a compilation of selected 
atomic data for resonant absorption lines 
at wavelengths longward of 911.753 {\AA} (the \ion{H}{1} Lyman limit), for  
key heavy elements (heavier than atomic number 5) of astrophysical interest. 
In particular, we focus on the transitions of those ions that have been 
observed in the Milky Way interstellar medium (ISM), the circumgalactic medium 
(CGM) of the Milky Way and/or other galaxies, and the intergalactic medium (IGM). 

  We provide wavelengths, oscillator strengths, associated accuracy grades, 
and references to the oscillator strength determinations. We also attempt to 
compare and assess the recent oscillator strength determinations. For about 
22\% of the lines that have updated oscillator strength values, the differences 
between the former values and the updated ones are $\gtrsim$~0.1 dex. 

  Our compilation will be a useful resource for absorption line studies of the 
ISM, as well as studies of the CGM and IGM traced by sight lines to quasars and 
gamma-ray bursts. Studies (including those enabled by future 
generations of extremely large telescopes) of absorption by galaxies against the 
light of background galaxies will also benefit from our compilation. \\

	{\it Supporting material:} machine readable table

\end{abstract}
	
\section{Introduction}
Atomic spectroscopy is fundamental to the study of a wide range of astrophysical 
environments. In the diffuse interstellar gas in the Milky Way, the atoms are 
often in the ground state, so that the resonant atomic transitions are of 
special interest. The vast majority of these atomic transitions lie in the 
ultraviolet. Space-based UV spectroscopy with a number of missions has made it 
possible to observe these interstellar transitions. Some of the earliest of 
these observations, carried out with the {\it Copernicus}  mission led to 
fundamental discoveries such as the hot halo gas of the Milky Way  [e.g., \cite{RSD73}; 
\cite{Y74}; \cite{SJ75}]. Subsequent 
missions such as the Far Ultraviolet Spectroscopic Explorer further extended the 
study of interstellar and intergalactic gas [e.g., \cite{M00}]. The several generations of UV 
spectrographs on the {\it Hubble Space Telescope} have vastly increased the number of 
Galactic as well as extragalactic sight lines probed for their neutral or 
ionized gas. For example, these observations uncovered the existence and 
properties of low-redshift Lyman-$\alpha$ forest clouds, as well as the covering 
fractions, element abundances, temperatures, and kinematics of the 
circumgalactic medium (CGM; e.g., \cite{MWS91}; \cite{BBB93}; 
\cite{SS96} and references therein; \cite{KFL05};  \cite{LHT13}; \cite{TTW13}; 
\cite{S15}; \cite{WPC16}). 
Naturally, these observations provide crucial 
constraints on models of galaxy evolution, including the effect of outflows 
and inflows. Furthermore, determinations of relative element abundances in the 
interstellar medium (ISM) are important to understanding dust depletions and 
thus, indirectly, the composition of dust grains [e.g. \cite{J09} and references therein]. 
Relative element abundances 
in distant galaxies offer crucial windows in understanding the cosmic evolution 
of dust, as well as the evolution of stellar nucleosynthetic processes. On a 
larger scale, observations of key ions in the intergalactic medium (IGM) offer 
rich insights into the physical conditions in the diffuse regions of the cosmic 
web, and the large-scale cosmic processes influencing it. 

In view of the sweeping consequences of atomic spectroscopy for understanding 
the evolution of galaxies, ISM, CGM, and IGM,  it is important to be able to 
derive accurate physical information from the spectra. This makes it essential 
to employ as accurate atomic data as possible in translating the spectroscopic 
measurements into determinations of physical quantities. 

Thanks to the extensive efforts of numerous theoretical and experimental 
physicists, many improvements in the atomic data relevant for astrophysical 
spectroscopy have been happening in recent years. However, knowledge of many of 
these improvements often does not trickle down to the community of observational 
spectroscopists rapidly enough. For example, in the CGM/IGM community, the most 
commonly used reference for atomic data, by far, is \cite{Morton03}. With the 
goal of making the latest improvements accessible to the community, here we 
present a compilation of oscillator strengths for key transitions, including 
updates made since 2003. We focus, in particular, on the ions that have been 
measured in ISM/CGM/IGM studies for selected elements ranging from C to Pb. 
For each of the selected elements, we list lines longward of the Hydrogen Lyman
limit at 911.753 {\AA}, since, in this wavelength region, the bound-free H 
absorption does not contribute much to the ISM/CGM/IGM opacity. Atomic data 
for absorption lines shorter than 911.753 {\AA} from the ground level can 
be found elsewhere, e.g. \cite{verner94} and \cite{Kal07}.

\renewcommand{\baselinestretch}{0.95}
\begin{deluxetable*}{llllllllllll}
\tabletypesize{\scriptsize}
\tablecolumns{12} 
\tablewidth{0pt}
\tablecaption{
Line Identifications, Observed Wavelengths $\lambda_{\rm vac}$ (\AA), 
Ritz Wavelengths $\lambda_{\rm Ritz}$ (\AA), Absorption Oscillator Strengths 
$f$, and Their Accuracy Grade for Key Transitions
}
\tablehead{ 
\colhead{$Z$} &
\colhead{Ion} &
\colhead{Lower level} &
\colhead{Upper level} &
\colhead{$\lambda_{\rm vac}$} & 
\colhead{$\lambda_{\rm Ritz}$} & 
\colhead{$g_{\rm l}$} &
\colhead{$g_{\rm u}$} &
\colhead{$f$}&
\colhead{$\log(gf)$} &
\colhead{Grade} &
\colhead{Source}
}
\startdata 
6 & C {\sc i} & 2s$^2$2p$^2$ $^3$P$_0$ & 2s$^2$2p3s $^3$P$^{\rm o}_1$  & 1656.928 & 1656.929& 1& 3& 1.43E$-$1& $-$0.845& A  & FF06 \\
6 & C {\sc i} & 2s$^2$2p$^2$ $^3$P$_0$ & 2s2p$^3$ $^3$D$^{\rm o}_1$    & 1560.310 & 1560.309& 1& 3& 7.16E$-$2& $-$1.145& A  & FF06 \\
6 & C {\sc i} & 2s$^2$2p$^2$ $^3$P$_0$ & 2s2p$^3$ $^3$P$^{\rm o}_1$    &  & 1328.834& 1& 3& 5.80E$-$2& $-$1.236& B  & FF06 \\
6 & C {\sc i} & 2s$^2$2p$^2$ $^3$P$_0$ & 2s$^2$2p4s $^3$P$^{\rm o}_1$  &  & 1280.135& 1& 3& 2.61E$-$2& $-$1.583& B+ & FF06 \\
6 & C {\sc i} & 2s$^2$2p$^2$ $^3$P$_0$ & 2s$^2$2p3d $^3$D$^{\rm o}_1$  & 1277.245 & 1277.245& 1& 3& 9.22E$-$2& $-$1.035& A  & FF06 \\
6 & C {\sc i} & 2s$^2$2p$^2$ $^3$P$_0$ & 2s$^2$2p5d $^3$D$^{\rm o}_1$  & 1157.910 & 1157.910& 1& 3& 2.12E$-$2& $-$1.674&    & ZF02 \\
6 & C {\sc ii}& 2s$^2$2p $^2$P$^{\rm o}_{1/2}$ &2s2p$^2$ $^2$D$_{3/2}$ & 1334.532 & 1334.532& 2& 4& 1.29E$-$1& $-$0.589&    & FFT04\\
6 & C {\sc ii}& 2s$^2$2p $^2$P$^{\rm o}_{1/2}$ &2s2p$^2$ $^2$S$_{1/2}$ & 1036.337 & 1036.337& 2& 2& 1.19E$-$1& $-$0.624&    & FFT04\\
\enddata                                                                              
                                                                                      
\tablecomments {
1. Table~\ref{tab_tran} is available in its entirety as a machine readable table.  A portion is shown here to illustrate its content.
 2. An estimated accuracy grade is listed for each oscillator strength where available, indicated by a 
code letter as given below:
AAA 	$\leq 0.3\%$;
$0.3\%<$ AA 	$\leq	1\%$;
$1\%<$ A+ 	$\leq	2\%$;
$2\%<$ A 		$\leq	3\%$;
$3\%<$ B+ 	$\leq	7\%$;
$7\%<$ B 		$\leq	10\%$;
$10\%<$ C+ 	$\leq	18\%$;
$18\%<$ C 	$\leq	25\%$;
$25\%<$ D+ 	$\leq	40\%$;
$40\%<$ D 	$\leq	50\%$;
E 	$> 50\%$.
}

\label{tab_tran}
\end{deluxetable*}

\section{Description of Compiled Data}

\subsection{General Terminology and Definitions}

Throughout this paper, we focus on the electric dipole (E1) transitions. The 
absorption line corresponds to the transition between the lower level l and 
the upper level u, with level energies $E_{\rm l}$ and $E_{\rm u}$, respectively. 
The statistical weights ($g = 2J +1$, $J$ being the total angular momentum of 
the state) of the lower and upper levels are denoted $g_{\rm l}$ and $g_{\rm u}$. 
In terms of the energy levels, the vacuum wavelength $\lambda_{\rm Ritz}$ of the 
transition is
\begin{equation}
\lambda_{\rm Ritz} =  hc/(E_{\rm u}-E_{\rm l}),
\end{equation}
where $h$ is Planck's constant, and $c$ is the speed of light. The Einstein 
transition probability (in s$^{-1}$) for spontaneous emission is denoted by 
$A_{\rm ul}$. The dimensionless absorption oscillator strength  $f_{\rm lu}$ 
is related to the E1 transition probability $A_{\rm ul}$ by
\begin{equation}
f_{\rm lu} = 1.49919 \times 10^{-16} \lambda_{\rm vac}^{2} g_{\rm u} A_{\rm ul}/g_{\rm l}
\end{equation}
where $\lambda_{\rm vac}$ is in \AA. The oscillator strengths compiled in this 
paper were obtained from either experimental techniques or theoretical 
calculations. In cases where oscillator strengths in both the length and 
velocity forms are available, we tabulate the length-form data because these are 
commonly used and more reliable.

\subsection{Tabulated Data}

In Table~\ref{tab_tran}, we present the data for each ion species separately. These
species and their ground levels are given in table subheaders.
Table~\ref{tab_tran} lists the following information for each E1 line of interest. 

\begin{enumerate}[noitemsep]

\item Nuclear charge $Z$.
\item Ion spectroscopic notation.
\item Lower level.
\item Upper level.
\item Vacuum rest wavelength $\lambda_{\rm vac}$ in {\AA} (where available).
\item Ritz wavelength $\lambda_{\rm Ritz}$ in {\AA}.
\item Lower level statistical weight $g_{\rm l}$.
\item Upper level statistical weight $g_{\rm u}$.
\item Absorption oscillator strength $f$.
\item Logarithm of weighted oscillator strength, $\log(gf)$.
\item Accuracy grade (uncertainty) of the oscillator strength $f$ according to the 
NIST atomic spectra database \citep{asd16}. 
\item $f$-value data source.

\end{enumerate}

\subsection{Selection rules}

We wish to be as consistent as possible in selecting the lines presented
in Table~\ref{tab_tran}. Hence we need to explain a selection procedure for
the tabulated lines. The following basic rules were followed for selecting data sources
in our tables or providing any additional information.

\begin{enumerate}[noitemsep]

\item
For the species range, we consider elements with $6 \leq Z \leq 30$ and several 
other elements observed in the ISM, such as Ga, Ge, Kr, and Pb. Usually we
present data for a few of the lowest ionization stages. The selected ions have been
observed in the ISM, CGM, and/or IGM.
\item
We tabulate lines with wavelengths $\lambda_{\rm vac} > 911.753$ \AA. These are
the vacuum wavelengths for all lines.
\item 
We give priority to the observed wavelength, which we call $\lambda_{\rm vac}$
over the Ritz wavelength $\lambda_{\rm Ritz}$. As a rule, the line wavelength
source is the NIST database  \citep{asd16}. We have represented the Ritz wavelengths from 
the NIST database to three decimal places.
\item
We tabulate absorption lines originating from the ground level only. We do not
tabulate lines originating from the excited levels of the ground configuration 
or the ground term even if their energies are just a few tenths of cm$^{-1}$, 
e.g. the \ion{C}{1}, \ion{C}{2}, and \ion{N}{2} ions.
\item
We tabulate only those lines that have $f \geq 0.001$. We present no more than 
three significant figures for the $f$-values because we suppose that is enough to 
reflect their real accuracy.
\item
Usually, we tabulate $f$-values from the newest sources giving priority to the
experimental data over the theoretical values. In the cases where new data
are not significantly different from the older data, we choose to rely on the 
older data preferring the most advanced theoretical methods for data production. 
Those special cases will be mentioned in  Sect~\ref{comments}.

\end{enumerate}

The accuracy grades of the tabulated oscillator strengths $f$ were either derived 
from the original data sources or we tabulated the grades given in the NIST database by  \cite{asd16}. 
For this reason, some of the lines do not have accuracy grades. In some cases, we were 
able to confidently assign an accuracy grade through careful comparative analysis. These
instances and their justification are specifically documented within Sect~\ref{comments}.

\section{Comments and Assessment of Improved Data}
\label{comments}
In Table~\ref{tab_tran}, wavelengths $\lambda_{\rm vac}$ are adopted from the
NIST database  \citep{asd16}. Theoretical $f$-values $f_{\rm theor}$ are corrected 
for the inaccuracy in calculated level energies (or wavelengths 
$\lambda_{\rm theor}$) by adjusting them according to the observed values 
$\lambda_{\rm vac}$, see \cite{stout}:
\begin{equation}
f_{\rm corr} = f_{\rm theor} \times (\lambda_{\rm theor} / \lambda_{\rm vac}).
\end{equation}

A significant part of the new transition data are utilized from the theoretical 
calculation performed by \cite{FFT04}, where the multiconfiguration Hartree-Fock 
(MCHF) method was applied to determine the transition data for the neutrals  
and ions starting with the beryllium isoelectronic sequence and finishing with
the argon isoelectronic sequence. The same MCHF approximation is applied to 
produce data for the sodium to argon isoelectronic sequences by \cite{FFT06}.
This method is a reliable one producing high-accuracy results, and it is 
difficult to exceed their accuracy when dealing with large amounts of species. 
We tabulate transition data from the above sources for the species up to argon
complemented with data from other sources for the lines involving higher 
excited levels with $n > 4$. Another source of oscillator strength $f$-values
for ions with $Z > 20$ is the data list of \cite{K16}. This is an online data list, 
which is continuously updated with new results. We give the preference to their 
newer data as opposed to the older values given in the previous versions of
this data list. We reference this source as \cite{K16}, though in many cases
data have been produced considerably earlier.

\subsection{Carbon species}
\label{C}

For neutral carbon and the species \ion{C}{2} and \ion{C}{3}, we tabulate
theoretical data from the calculations of Froese Fischer and co-workers 
published in \cite{FF06}, \cite{ZFF02}, and \cite{FFT04}. 
These are very reliable data sources providing highly accurate radiative 
transition data. For two lines in the lithium-like \ion{C}{4}, which is 
outside the scope of the above-mentioned papers, 
we adopt data from \cite{YTD98} with an accuracy grade of A. We performed
an additional check for these two lines and calculated transition data using
our own codes utilizing the Hartree-Fock (HF) and quasirelativistic (QR) 
approaches on a very extensive configuration-interaction (CI) wavefunction 
basis recently described by \cite{rk_zn2}. Our results confirm the high accuracy 
of data from \cite{YTD98}; our calculated $f$-values agree within a few tenths 
of a percent.

\subsection{Nitrogen species}
\label{N}

Oscillator strengths are taken from \cite{FFT04} for the species \ion{N}{1},
\ion{N}{2}, and \ion{N}{3}. The NIST database assigns them accuracy grades no 
worse than B. There is just one line in our investigated wavelength range in 
the \ion{N}{4} ion. It originates from the excited level 1s$^2$2s2p $^3$P$_1$. 
However, it is rather weak because it constitutes a spin-changing E1 transition. 
For the lithium-like ion \ion{N}{5}, we adopt data from \cite{PSS88}. For the
\ion{N}{5} lines, we performed an accuracy check because these data are
coming from non-relativistic calculations. Our results, both in the length and
velocity forms of the E1 transition operator, agree between themselves within
$2\%$ and do not deviate more than $0.5\%$ from the results of \cite{PSS88}.
This confirms a high-accuracy grade of A assigned to oscillator strengths
of the tabulated absorption lines in the \ion{N}{5} ion.

\subsection{Oxygen species}
\label{O}

For the \ion{O}{1} ion, data from \cite{FFT04} are adopted for the lines 
connecting the excited states 2s$^2$2p$^3$3s, 3d, and 4s. For other lines not covered
by \cite{FFT04}, we use data from \cite{BZ91} and \cite{HBG91}. The latter is the same source used
by the NIST database, and the later investigation of \cite{T09} just confirms the
reliability and high accuracy of their data. Transition data from \cite{T09}
agree with oscillator strengths $f$ of \cite{HBG91} within $10\%$.
We also tabulate oscillator strengths for two 2s -- 2p transition lines in
Li-like \ion{O}{6} determined by \cite{PSS88}. Likewise to the situation
described in Sect~\ref{N}, we have performed our own calculation and can confirm
a high-accuracy grade assigned to these data by the NIST team.

\subsection{Sodium species}
\label{Na}

The oscillator strength $f$-values are taken from \cite{FFT06} for the lines
3s -- $n$p, ($n=3,4$), and we retain the same accuracy grades as given by NIST.
For the line 3s -- 5p, we adopt the oscillator strength value determined by applying
the same MCHF approach $B$-spline method with non-orthogonal radial orbitals by 
\cite{FF02}. It was proved for neutral carbon by \cite{ZFF02} that such an
approach produces reliable results for the high-$nl$ lines.

\subsection{Magnesium species}
\label{Mg}
For the lines representing resonance transitions from the ground 3s$^2$ state 
to the excited 3s3p and 3s4p states of the \ion{Mg}{1} ion, we use data from
\cite{FFT06} with accuracy grades of A and B+. For the transition to the 3s5p 
levels, more recent oscillator strength $f$ data produced using
a $B$-splines method within MCHF by \cite{ZBG09} are available, 
whereas for the lines representing transitions to highly excited  3s$n$p levels 
($6 \leq n \leq 8)$, we adopt data of \cite{CT90} because more recent data do not
exist.

Oscillator strength values for \ion{Mg}{2} are taken from \cite{FFT06}. 
The accuracy of the data for the 3s -- 3p lines has a very high A+ grade. Meanwhile,
though the lines 3s -- $n$p ($n > 3$) are in our investigated wavelength range
($\lambda_{\rm vac} > 911.753$), only the transition to the 2p$^6$5p 
$^2$P$^{\rm o}_{3/2}$ level has $f \geq 0.001$ satisfying our selection criteria.
Other lines fall short of that criteria, though their fine-structure level
unresolved oscillator strengths are within that range.

\subsection{Aluminum species}
\label{Al}

We tabulate oscillator strength $f$-values for three ions, \ion{Al}{1}, 
\ion{Al}{2}, and \ion{Al}{3}. Here for the lines connecting the ground state 
with the lower $n = 3, 4$ levels, the theoretical data of \cite{FFT06} are 
adopted. They have accuracy grades of B+ and A+. For the higher $nl$ levels,
we list theoretical data from \cite{MEL95}, which were determined by 
non-relativistic R-matrix calculations. Furthermore, their data for fine-structure
levels were derived from the multiplet values assuming a pure $LS$-coupling.
Those data usually have rather poor accuracy grades, thus further data 
improvement can rectify this situation. For three lines of \ion{Al}{1},
we adopt experimental data of \cite{VBF02} and \cite{DVD90} and assign an
accuracy grade of B.

\subsection{Silicon species}
\label{Si}

For the low-lying states of the 3s$^2$3p4s, 3s3p$^3$, 3s$^2$3p3d 
odd-parity configurations of \ion{Si}{1}, we list $f$-values adopted from \cite{FFT06}. As we 
have mentioned before, these theoretical data are of high quality and 
reliability. It is necessary to admit here that these values differ noticeably  
from the previous oscillator strength values tabulated by \cite{Morton03}. 
The case of the lines involving higher levels is more complicated. Some lines 
falling into our wavelength selection region are absent from the NIST database, though their corresponding level energies are presented there, e.g. the lines at 
$\lambda_{\rm Ritz}= 1568.617$ {\AA}, $\lambda_{\rm Ritz}= 1763.661$ {\AA}, 
and $\lambda_{\rm Ritz}= 1873.103$ {\AA}. The oscillator strength value for 
the line at $\lambda_{\rm Ritz}= 1589.173$ {\AA} is 
derived from the non-relativistic R-matrix calculations of collision data for 
several Si-like ions of \cite{np93}, which we consider to be less accurate 
compared to the pure atomic structure calculations or the experimental results. 
The E1 transition data for the latter line are listed in \cite{FFT06}, though 
their $\log(gf)$ value of $-2.16$ differs noticeably from the experimental 
$\log(gf) = -2.57$ value from \cite{SHT87}. Additional experimental 
data for higher $nl$ levels comes from the measurements of \cite{SHT87} and 
from a critical compilation of \cite{KP08}.

\cite{BQP09} have produced a benchmark data set for the \ion{Si}{2} ion by
utilizing several theoretical approaches and experimental data to determine
reliable recommended absorption oscillator strengths for the levels of the
3s3p$^2$, 3s$^2$3d, and 3s$^2$4s configurations. For the higher level of 
3s$^2$4d at 989.873 {\AA}, we list the experimental $f$-value of \cite{CS74}, 
which is close to the theoretical value $f=0.1849$ from \cite{CM00}, which also
serves as a source for the 3s$^2$5s level data. For the ions \ion{Si}{3} and
\ion{Si}{4}, we list oscillator strengths from \cite{FFT06}.

\subsection{Phosphorus species}
\label{P}

For the lines connecting the ground state 3s$^2$3p$^3$ $^4$S$^{\rm o}_{3/2}$ with
the levels of $^4{\rm P}$ term of the excited configuration 
3s$^2$3p$^2$($^3{\rm P}$)4s of \ion{P}{1}, we adopt oscillator strengths from 
\cite{FFT06}. Unfortunately, that work does not list data for the transitions to
the 3s$^2$3p$^2$($^3{\rm P}$)3d $^4{\rm P}$ levels. Therefore, for these levels,
we resort to older theoretical data from \cite{F86}. The $f$-values for the 
lines at $\lambda\lambda  1679.695, 1674.591, 1671.680$ given by \cite{FFT06}
differ significantly, at least by two orders of magnitude from the previous data
of \cite{F86}. Since the relative intensities for these three 3s -- 3p lines 
given by the NIST database are similar to those of 3p -- 4s lines at 
$\lambda\lambda 1787.656, 1782.838, 1774.951$, we believe the data from 
\cite{F86} given its D accuracy rating can be utilized for neutral 
 phosphorus.

For the \ion{P}{2} and \ion{P}{3} lines, we list data from \cite{FFT06}. Five 
more lines below 1153 {\AA} that are unlisted in the NIST database are entered 
in Table\,\ref{tab_tran} for \ion{P}{2}. These lines originate from the 
absorption transitions from the ground level to the levels 3p4s and 3p3d.
We have derived their Ritz wavelengths using level energies listed by 
 \cite{asd16}. The same process was performed for three \ion{P}{3} lines 
below 999 \AA. The source of the transition data chosen for the magnesium-like 
\ion{P}{4} and sodium-like \ion{P}{5} is \cite{FFT06}.

\subsection{Sulfur species}
\label{S}

The main source of the data for the neutral sulfur lines connecting 
the ground level 3s$^2$3p$^4$\,$^3$P$_2$ with the levels of the excited 
3d, 4d, 4s, 5s, and 6s configurations are the theoretical results of \cite{DH08}.
As in \cite{DH06}, the data from \cite{DH08} agree very well
with most lines from the results of the $B$-spline calculations by
\cite{ZB06}. They are also close to the data produced by \cite{FFT06}. 
Thus one has to be assured of high quality and reliability of the listed 
$f$-values. We include the line at $\lambda=1474.5715$\,{\AA} even though its $f$-value
is lower than our selection criterion $f \le 0.001$ as other theoretical 
predictions put it above this criterion. The data from higher configurations 
with the valence 5d, 6d, 7s, and 8s electrons are taken from \cite{BGF98}. 
That set of $f$-values was deduced from  a combination of laser lifetime 
measurements and theoretical branching ratios. For the lines with 
$\lambda < 1241$\,{\AA} that connect the ground state with the upper levels located
above the first ionization limit at $83,559.1$\,cm$^{-1}$, we choose oscillator
strengths from \cite{DH08}, which have an accuracy grade of C.

For the \ion{S}{2} lines, we list a data set from \cite{rk2014}. The authors 
concluded that their data are in good agreement with other theoretical data 
sets of \cite{FFT06} and \cite{TZ10}. For the \ion{S}{3}, \ion{S}{4}, and 
\ion{S}{6} lines, we list data from \cite{FFT06}.

\subsection{Chlorine species}
\label{Cl}

The data for the \ion{Cl}{1} lines are taken from the recent theoretical 
calculations of \cite{OH13}. Their data accuracy grade is C+. 
For the two lines in \ion{Cl}{2} and three lines of \ion{Cl}{3}, we assume 
that the experimental data from \cite{SFB05} are the best source. The accuracy 
of their $f$-values is within 5\% (or an accuracy grade of B+). We list an 
oscillator strength from \cite{FFT06} for a single line of \ion{Cl}{4}.

\subsection{Argon species}
\label{Ar}

For the lines of \ion{Ar}{1} and \ion{Ar}{2}, we recommend using oscillator
strengths from \cite{FFT06}. One can safely assign the C+ accuracy grade to
these data.

\subsection{Potassium species}
\label{K}

For the spectra of neutral \ion{K}{1} we list four lines corresponding to the
resonance 4s -- 4p and 4s -- 5p transitions in the valence shell. The lines to the
higher $n$p levels are relatively weak and do not fall into our desired
$f$-value range. The parameters of lines caused by the 4s -- 4p transitions are
determined by \cite{WWW97}. Their data agree very well with other 
high-accuracy measurements. They evaluate the accuracy of $f$-values as
0.2\% (AAA accuracy grade). Data for the 4s -- 5p lines are from the measurements
of \cite{SK85}. Although their accuracy grade is unlisted, by comparing these
data with the results of elaborated relativistic calculations of
\cite{Migdalek98}, we confidently assign them a grade of A.

\subsection{Calcium species}
\label{Ca}

The data for the $\lambda$4227.92 line are taken from the high-accuracy 
photoassociative measurements of \cite{ZBR00}, which produces reduced 
uncertainties compared to previous measurements. Their data have confirmed
previous level-crossing measurements of \cite{KS74}. Recent theoretical values 
of \cite{FFT03} are also close (within error bars) to the above experimental data. 
The data for other resonance lines of neutral calcium are measured by 
\cite{PRT76}. They used the hook method  to determine the $gf$-values of the 
4s$^2$ -- 4s$n$p absorption lines. These data for most lines agree within 
5\% with other measurements, e.g. \cite{OP61}, \cite{S63}, and \cite{Mit75}. 
The $\lambda$2722.450 line is the only exception, and here the differences are 
much higher. Even though this line has an $f$-value below our cut-off level, 
we list it in our table. The list for \ion{Ca}{1} contains lines up to 
3p$^6$4s13p, while the transitions for higher Rydberg levels fall below our 
cut-off criterion of $f \ge 0.001$. 

For the \ion{Ca}{2} ion
lines, we recommend recently determined oscillator strengths from \cite{SS11}.
Their calculations used a high-precision relativistic method, where all 
single and double, and partially triple excitations of Dirac-Fock wavefunctions 
are included to all orders of perturbation theory. The authors estimate the
relative uncertainties of their calculated oscillator strengths to be 0.9\%.

\subsection{Titanium species}
\label{Ti}

The majority of the data for the \ion{Ti}{1} lines come from the measurements 
of \cite{LGW13}. They have used the previously measured radiative lifetimes 
combined with the branching fractions to yield absolute oscillator strengths. 
The data agree closely with the NIST data; therefore we consider it appropriate 
to assign them respective accuracy grades, which are no worse than B+. 
Some lines are missing in \cite{LGW13}. For those lines, we recommend 
the high-quality experimental data from \cite{BLN06} or data from the earlier 
experiment of \cite{SK78}.

For the \ion{Ti}{2} ion lines, we list data from the recent paper of 
\cite{LHE16}. The authors measured radiative level lifetimes and used the 
pseudo-relativistic HF method to determine oscillator strengths
and theoretical lifetimes for the measured levels. Their data agree within
10\% with the earlier theoretical data of \cite{RED14} and those from 
\cite{PTP01}. For the \ion{Ti}{3} transition data, we adopt the theoretical 
results of \cite{RU97}. They performed an orthogonal operator calculation for 
the electric dipole transition integrals by means of the multiconfiguration 
Dirac-Fock method including core polarization.

\subsection{Chromium species}
\label{Cr}

For the lines connecting the ground state of \ion{Cr}{1} with the $n=4$ levels,
the data chosen were produced by \cite{SLS07}. The authors used branching
fraction measurements from Fourier transform spectra in conjunction with 
radiative lifetimes to determine the transition probabilities. These data are
assigned an accuracy grade of B. For the transitions to the 3d$^5$5p,6s levels, we 
list data from \cite{MFW88}, which are given accuracy grades of C.

For the three lines above 2050 {\AA} in \ion{Cr}{2}, the recommended tabulated 
data are from 
the experimental work of \cite{NLL06}. The accuracy grade for their data is B+. 
For other lines of \ion{Cr}{2} we tabulate theoretical data from \cite{RU98},
where oscillator strengths have been determined using the orthogonal operator
technique. Five listed lines of \ion{Cr}{3} are taken from \cite{K16}.

\subsection{Manganese species}
\label{Mn}

For the manganese lines, we adopt the experimental oscillator strength values 
from \cite{BXP05, BPN11}. Their accuracy grade is B+ for the lines with
$\lambda >4000$\,{\AA}, and C+ for the lines below 3000 {\AA}. Unfortunately,
oscillator strengths for most \ion{Mn}{1} lines listed in our table were not 
measured in the above experiments. For these lines, we tabulate older data 
from \cite{MFW88} or data from \cite{K16}.

New data for some selected \ion{Mn}{2} lines in Table~1 are listed from 
\cite{DHL11} who experimentally measure radiative 
lifetimes and branching fractions to derive transition probabilities and
oscillator strengths. Other data come either from similar experimental work 
of \cite{KG00} or from the configuration-interaction calculations of 
\cite{TH05}.

\subsection{Iron species}
\label{Fe}

Lines for three iron ions are listed in Table\,1. For lines with $\lambda > 2913$\,
{\AA} our oscillator strength source is \cite{BIP79}, who
measured absorption oscillator strengths and reported an accuracy of 0.5\%. 
NIST assigns an accuracy grade of A for most of those lines. For the lines 
below 2913\,{\AA}, our data come from the measurements of \cite{OWL91}. They 
employed time-resolved laser-induced fluorescence to measure radiative 
lifetimes and derived oscillator strengths by measuring branching
fractions. Their accuracy grades are slightly worse compared to those of 
\cite{BIP79}. For the remaining lines below 2260\,{\AA}, we list the data of
\cite{BH73}, who used the hook method to determine oscillator strengths in 
\ion{Fe}{1}. Although the accuracy grade of their data is not as high, these
oscillator strengths are substantially reliable.

Our main source for the \ion{Fe}{2} lines above 2000\,{\AA} is the experimental
data from \cite{BMW96}. The authors have measured branching ratios with a 
Fourier transform spectrometer and with a high-resolution grating spectrometer. 
The resulting measurements were used to derive transition probabilities for 56 
lines. Another group of listed lines was theoretically studied by \cite{RU98}. 
Their accuracy grade is lower. These two sources are complemented by 
experimental oscillator strengths from \cite{PJS01} and the critical 
compilation of \cite{FW06}. For the \ion{Fe}{3} ion line at 1122.5 {\AA}, we 
tabulate the theoretical oscillator strength from \cite{DH09}.

\subsection{Cobalt species}
\label{Co}

Oscillator strengths for all but one of the \ion{Co}{1} lines are taken from 
the recent  measurements of \cite{LSC15}. They derived oscillator strengths 
from experimental branching fractions combined with radiative lifetimes
from laser-induced fluorescence measurements. The NIST database assigns 
accuracy grades of B or B+ for their data. A rather weak line at $\lambda$2436 
is taken from \cite{CSS82}, where absolute oscillator strengths were determined
using the hook method. Data for the \ion{Co}{2} lines come from the calculations 
using the orthogonal operator technique by \cite{RPU98} and the measurements 
of \cite{MCL98} and \cite{MLZ98}. In general, the data for \ion{Co}{2} have relatively low 
accuracy grades. For the lines of \ion{Co}{3}, we tabulate oscillator strengths
from the Kurucz online data list \citep{K16}.

\subsection{Nickel species}
\label{Ni}

For the \ion{Ni}{1} lines, we list oscillator strengths from new measurements
of \cite{WLS14}, where $f$-values were determined by combining measured branching
fractions and radiative lifetimes. Most of these data are assigned high-accuracy grades (from B+ to B). Missing data are covered by the measured 
oscillator strengths obtained by the hook and absorption methods from
\cite{HS80}. Their data accuracy evaluation is lower compared to that of
\cite{WLS14}. NIST lists only a single line for transitions of \ion{Ni}{2}.
We tabulated oscillator strengths from the recent CI calculation of 
\cite{CHR16}. The authors state that it is difficult to provide a measure of 
the uncertainties for the large-scale CI calculations, which would cover all 
transitions. Nevertheless, they consider that their data are the best currently
available and meet the accuracy demands for astrophysical applications. 
We do not assign any accuracy grades to these data, though they can be 
considered as having an accuracy grade of B.

\subsection{Copper species}
\label{Cu}

The source of oscillator strength data for the resonance 4s -- 4p lines in 
\ion{Cu}{1} is a critical compilation of oscillator strengths for neutral ions 
by \cite{D95}. The data for the transitions 3d -- 4p and 4s -- 5p are adopted 
from \cite{HM78}. The oscillator strengths were determined from atomic 
absorption measurements of the radiation emitted from a copper hollow-cathode 
lamp. The oscillator strengths for the remaining lines 4s -- $n$p $(n \geq 6)$ 
comes from the calculation of \cite{CAE15}.

Three lines listed for \ion{Cu}{2} originate from the transitions involving the
ground level 3p$^6$3d$^{10}$\,$^1$S$_0$ and the levels with $J=1$ from the 
configuration 3d$^9$4p. Their data source is either the theoretical data from
\cite{DHB99} or from the beam-foil experimental data of \cite{BFI09}, which 
were given an accuracy grade of B.

\subsection{Zinc species}
\label{Zn}

The oscillator strength for the 4s$^2$\,$^1$S$_0$ -- 4s4p$^1$P$^{\rm o}_1$ line of
\ion{Zn}{1} is taken from the Hanle-effect experiments of \cite{KT76}, which
have confirmed the earlier level-crossing technique results of \cite{LZG64}.
Later beam-foil spectroscopy measurements of \cite{MCH79} give very similar
$f$-values ($f=1.55 \pm 0.08$). These measurements are consistent with recent 
theoretical data, e.g. \cite{FFZ07} and \cite{GM06}. For the $\lambda1589$ 
line, we tabulate the oscillator strength from the line list of \cite{K16}.
The NIST database lists a rather strong line at $\lambda1109.1$ with no 
transition identification or Ritz wavelength. Analyzing the level list, we can
assume that this line originates from the transition 3d -- 4p with 
$\lambda_{\rm Ritz}=1108.316$ {\AA}, though this wavelength noticeably 
differs from the measured one. This can be explained by the fact that the level
3d$^9$4s$^2$4p\,$^3$P$^{\rm o}_1$ lies in the continuum, above the 
ionization limit.

Oscillator strengths for \ion{Zn}{2} are taken from the quasirelativistic 
calculations of \cite{rk_zn2}. Their data are consistent and agree with other 
theoretical data within a range of 10\%.

\subsection{Gallium species}
\label{Ga}

Tabulated data for neutral gallium lines are from a new compilation of
\cite{SRK07} with an unlisted line for the 4p$_{1/2}$ -- 7d$_{3/2}$ transition
data taken from the relativistic many-body perturbation theory calculation of 
\cite{SCS06}. These two data sources tabulate oscillator strengths differing
by approximately 10\% for most of the listed lines. 
For the $\lambda$1414 line of \ion{Ga}{2}, we tabulate an oscillator strength
from the low-energy beam-foil measurements of \cite{AEP79}, which have 8\%
measurement errors (i.e., an accuracy grade of B). Later theoretical data 
confirm the reliability of the listed experimental data, see, e.g. \cite{MEH05}, 
\cite{Jon06}, and \cite{CC14}. Oscillator strength values for two lines of the
\ion{Ga}{3} ion are from the compilation of \cite{SRK07}.

\subsection{Germanium species}
\label{Ge}

There are no new reliable original oscillator strength data sources for the 
 germanium species following \cite{Morton03}, except for the compilation 
of \cite{FW05}. For the \ion{Ge}{1} lines, we tabulate experimental oscillator 
strengths of \cite{LNP99}. Their experiment measured natural level radiative 
lifetimes by employing time-resolved UV laser-induced fluorescence from a 
laser-produced plasma and determined branching fractions by an inductively 
coupled plasma emission spectrometry technique. The derived oscillator 
strengths have accuracy grades of B or B+. For the \ion{Ge}{2} lines, we list 
theoretical $f$-values from \cite{BMQ98}. For the 
\ion{Ge}{3} ion, we list two 4s$^2$ -- 4s4p transition lines. The $^3$P$^{\rm o}_1$ 
spin-forbidden line oscillator strength is tabulated from the empirical 
predictions of \cite{C92}. The resonance line $f$-value for the $^1$P$^{\rm o}_1$ 
level was experimentally determined by \cite{AEP79} utilizing beam-foil 
spectroscopy. The oscillator strengths for \ion{Ge}{4} are adopted from
the beam-foil measurements of \cite{PBI81}.

\subsection{Krypton species}
\label{Kr}

For the \ion{Kr}{1} lines we tabulate experimental oscillator strengths from 
\cite{CCG92}. They measured photoabsorption $f$-values using the dipole ($e,e$) 
method. The oscillator strengths have accuracy grades of B or B+. 
The reliability of their data was confirmed by recent $B$-spline calculations 
of \cite{ZB09}. The oscillator strengths for \ion{Kr}{2} were derived from the 
selective pulsed monochromatized synchrotron radiation experiment of 
\cite{LLV99}, where lifetimes of the 4s4p$^6$\,$^2$S$_{1/2}$ states were 
measured. For the \ion{Kr}{6} line, we list data from MCHF relativistic calculations of \cite{PRB96}, where adjustments were 
made to the electrostatic parameters in order to improve theoretical level 
energies and $gf$-values.

\subsection{Lead species}
\label{Pb}

We list data for the \ion{Pb}{2} lines. The oscillator strength for the 
$\lambda1682$ line is adopted from \cite{QBP07}, where transition probabilities
were calculated in a relativistic multiconfiguration Dirac-Fock approach
and the accuracy of the results was assessed by a new experimental determination
of the radiative lifetime for the 7s$_{1/2}$ level. Other listed data  are from 
\cite{SSJ05}. The authors obtained radiative transition data using relativistic 
many-body perturbation theory. The remaining data for the 10s$_{1/2}$ and
9d$_{3/2}$ levels are adopted from the ab initio relativistic Hartree-Fock
calculations of \cite{CAM01}.

\section{Discussion}

A complete version of Table~\ref{tab_tran} is available online in a 
machine-readable version. Table~\ref{tab_tran} can also be 
obtained on request as a formatted PDF table from the authors 
(\texttt{Romualdas.Kisielius@tfai.vu.lt, fcashman@email.sc.edu}). In the printed version of 
this paper, we list just a few lines from the carbon species to facilitate 
understanding of the form and contents of our data list. 

In Table~\ref{tab_biblio}, we explain the source abbreviations used in 
Table~\ref{tab_tran}. For easier guidance, we also tabulate the ion list
from which oscillator strength values $f$ were sourced.

\begin{figure}[!th]
\includegraphics[width=\linewidth]{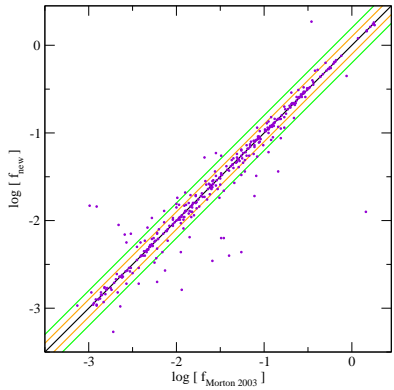}
\caption{ 
Comparison of the updated and former values of the oscillator strengths. 
The black line denotes a line of unit slope and zero intercept. 
The orange and green lines denote deviations of 
$\pm 0.1$ dex and $\pm 0.2$ dex, respectively, from the black line. 
\label{fig:fvaluecomp}
}
\end{figure}

Table~\ref{tab_tran} lists 576 transitions. For 400 of these transitions, 
we have listed updated oscillator strength determinations. Of these, 60 
transitions, though listed either in \cite{Morton03} or in the NIST database
 \cite{asd16}, previously had no oscillator strength value reported. 
Figure~\ref{fig:fvaluecomp} shows a comparison of the updated and former 
oscillator strengths. Table~\ref{tab:dex} compares updated oscillator strength 
values to their former values for 340 lines. For 171 of these transitions, 
the new $f$-values are smaller than the old values, while for 157 transitions, 
the new $f$-values are larger. The differences from the old values 
are usually smaller for the stronger transitions with log $f$ $\gtrsim -0.5$. 
The differences are $\ge 0.2$ dex for $\approx 12\%$ of the lines with changed $f$-values, and 
$\gtrsim 0.1$ dex for $\approx 22 \%$ of the lines. 

\begin{figure}[!th]
\includegraphics[width=\linewidth]{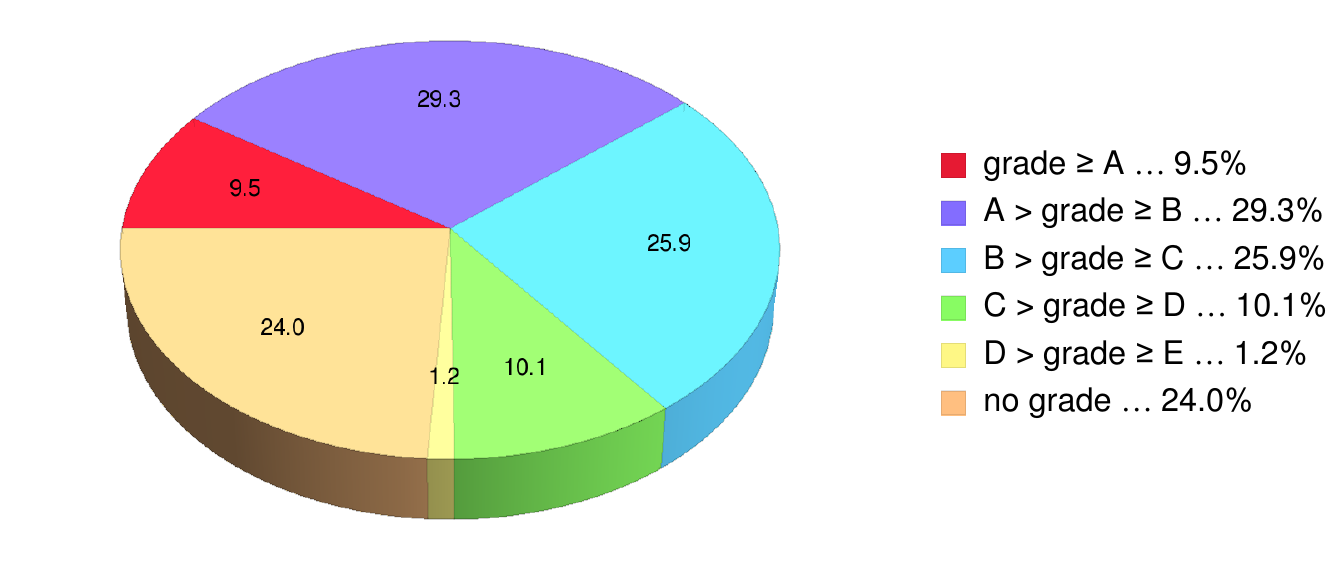}
\caption{Statistical breakdown of accuracy grades of the 576 transitions 
in Table~\ref{tab_tran}. An estimated accuracy grade is listed for each 
oscillator strength where available, indicated by the following code letters:
 grade $\ge$ A :	accuracy $\le	3\%$;
A$>$ grade $\ge$B: $3\%<$ accuracy $\le  10\%$; 
 B$>$ grade $\ge$C: $10\%<$ accuracy $\le	25\%$;
C$> $grade $\ge$D: $25\%<$ accuracy $\le	50\%$;
D$>$ grade $\ge$ E : accuracy $> 50\%$. 
\label{fig:pie}
}
\end{figure}

A breakdown of the accuracy grades for all 576 oscillator strengths is given in 
Fig.~\ref{fig:pie}. Approximately 37\% of the oscillator strengths have an 
accuracy grade worse than 10\%, while approximately 11\% of the oscillator 
strengths have an accuracy grade worse than 25\%. Figure~\ref{fig:pie} 
highlights the need for obtaining more accurate oscillator strength values for 
these, as well as for obtaining the accuracy grades for the 24\% of 
the oscillator strengths that are currently without a grade. 


\renewcommand{\baselinestretch}{0.95}
\noindent
\begin{deluxetable}{lll}
\tabletypesize{\scriptsize}
\tablecolumns{3} 
\tablewidth{0pt}
\tablecaption{
References Abbreviated in Table 1} 
\tablehead{ 
\colhead{Abbreviation} &
\colhead{Citation} & 
\colhead{Ion}
\label{tab_biblio}
}
\startdata 
AEP79 & Andersen et al. 1979 & Ga {\sc ii}, Ge {\sc iii} \\
BFI09 & Brown et al. 2009 & Cu {\sc ii} \\
BGF98 & Bi{\'e}mont et al. 1998 & S {\sc i} \\
BH73 & Banfield \& Huber 1973 & Fe {\sc i} \\
BIP79 & Blackwell et al. 1979 & Fe {\sc i} \\
BLN06 & Blackwell-Whitehead et al. 2006 & Ti {\sc i} \\
BMQ98 & Bi{\'e}mont et al. 1998 & Ge {\sc ii} \\
BMW96 & Bergeson et al. 1996 & Fe {\sc ii} \\
BPN11 & Blackwell-Whitehead et al. 2011 & Mn {\sc I} \\
BQP09 & Bautista et al. 2009 & Si {\sc ii} \\
BXP05 & Blackwell-Whitehead et al. 2005 & Mn {\sc i} \\
BZ91 & Butler \& Zeippen 1991 & O {\sc i} \\
C92 & Curtis 1992 & Ge {\sc iii} \\
CAE15 & {\c C}elik et al. 2015 & Cu {\sc i} \\
CAM01 & Col{\'o}n \& Alonso-Medina 2001 & Pb {\sc ii} \\
CCG92 & Chan et al. 1992 & Kr {\sc i} \\
CHR16 & Cassidy et al. 2016 & Ni {\sc ii} \\
CM00 & Charro \& Mart{\'i}n 2000 & Si {\sc ii} \\
CS74 & Curtis \& Smith 1974 & Si {\sc ii} \\
CSS82 & Cardon et al. 1982 & Co {\sc i} \\
CT90 & Chang \& Tang 1990 & Mg {\sc i} \\
D95 & Doidge 1995 & Cu {\sc i} \\
DH01 & Donnelly \& Hibbert 2001 & Fe {\sc ii} \\
DH08 & Deb \& Hibbert 2008 & S {\sc i} \\
DH09 & Deb \& Hibbert 2009 & Fe {\sc iii} \\
DHB99 & Donnelly et al. 1999 & Cu {\sc ii} \\
DHL11 & Den Hartog et al. 2011 & Mn {\sc ii} \\
DVD90 & Davidson et al. 1990 & Al {\sc i} \\
F86 & Fawcett 1986 & P {\sc i} \\
FF02 & Froese Fischer 2002 & Na {\sc i} \\
FF06 & Froese Fischer 2006 & C {\sc i} \\
FFT04 & Froese Fischer \& Tachiev 2004 & C {\sc ii}; {\sc iii}, N {\sc i}, {\sc ii}, {\sc iii}; \\
& & O {\sc i} \\
FFT06 & Froese Fischer et al. 2006 & Na {\sc i}; Mg {\sc i}, {\sc ii}; \\ 
& & Al {\sc i}, {\sc ii}, {\sc iii};  \\
& & Si {\sc i}, {\sc iii}, {\sc iv}; \\
& & P {\sc i}, {\sc ii}, {\sc iii}, {\sc iv}, {\sc v}; \\
& & S {\sc iii}, {\sc iv}, {\sc vi}; Cl {\sc iv}; \\
& & Ar {\sc i}, {\sc ii} \\
FW06 & Fuhr \& Wiese 2006 & Fe {\sc ii} \\
HBG91 & Hibbert et al. 1991 & O {\sc i} \\
HM78 & Hannaford \& McDonald 1978 & Cu {\sc i} \\
HS80 & Huber \& Sandeman 1980 & Ni {\sc i} \\
K16 & Kurucz 2016 & Fe {\sc ii}, Cr {\sc iii}, Mn {\sc i}, \\
& & Co {\sc iii}, Zn {\sc i} \\
KG00 & Kling \& Griesmann 2000 & Mn {\sc ii} \\
KKF14 & Kisielius et al. 2014 & S {\sc ii} \\
KKF15 & Kisielius et al. 2015 & Zn {\sc ii} \\
KP08 & Kelleher \& Podobedova 2008 & Si {\sc i} \\
KT76 & Kowalski \& Tr{\"a}ger 1976 & Zn {\sc i} \\
LGW13 & Lawler et al. 2013 & Ti {\sc i} \\
LHE16 & Lundberg et al. 2016 & Ti {\sc ii} \\ 
LLV99 & Lauer et al. 1998 & Kr {\sc ii} \\
LNP99 & Li et al. 1999 & Ge {\sc i} \\
LSC15 & Lawler et al. 2015 & Co {\sc i} \\
MCL98 & Mullman et al. 1998 & Co {\sc ii} \\
MEL95 & Mendoza et al. 1995 & Al {\sc i} \\
MFW88 & Martin et al. 1988 & Cr {\sc i}, Mn {\sc i} \\
MLZ98 & Mullman et al. 1998 & Co {\sc ii} \\
NLL06 & Nilsson et al. 2006 & Cr {\sc ii} \\                                                                 
NP93 & Nahar \& Pradhan 1993 & Si {\sc i} \\                                                                  
OH13 & Oliver \& Hibbert 2013 & Cl {\sc i} \\                                                                  
OWL91 & O'Brian et al. 1991 & Fe {\sc i} \\
PBI81 & Pinnington et al. 1981 & Ge {\sc iv} \\                                                                 
PJS01 & Pickering et al. 2001 & Fe {\sc ii} \\                                                                 
PRB96 & Pagan et al. 1996 & Kr {\sc vi} \\
PRT76 & Parkinson et al. 1976 & Ca {\sc i} \\
PSS88 & Peach et al. 1988 & N {\sc v}, O {\sc vi} \\                                                             
QBP07 & Quinet et al. 2007 & Pb {\sc ii} \\                                                                 
RPU98 & Raassen et al. 1998 & Co {\sc ii} \\
RU97 & Raassen \& Uylings 1997 & Ti {\sc iii} \\
RU98 & Raassen \& Uylings 1998 & Cr {\sc ii}, Fe {\sc ii} \\
SCS06 & Safronova et al. (2006) & Ga {\sc i} \\
SFB05 & Schectman et al. 2005 & Cl {\sc ii}, {\sc iii} \\
SHT87 & Smith et al. 1987 & Si {\sc i} \\
SK78 & Smith \& K{\"u}hne 1978 & Ti {\sc i} \\
SK85 & Shabanova \& Khlyustalov 1985 & K {\sc i} \\
SLS07 & Sobeck et al. 2007 & Cr {\sc i} \\
SMK98 & Siegel et al. 1998 & Mg {\sc ii} \\
SRK07 & Shirai et al. 2007 & Ga {\sc i}, {\sc iii} \\ 
SS11 & Safronova \& Safronova 2011 & Ca {\sc ii} \\
SSJ05 & Safronova et al. 2005 & Pb {\sc ii} \\
TH05 & Toner \& Hibbert 2005 & Mn {\sc ii} \\
VBF02 & Vujnovi{\'c} et al. 2002 & Al {\sc i} \\
WLS14 & Wood et al. 2014 & Ni {\sc i} \\
WWW97 & Wang et al. 1997 & K {\sc i} \\
YTD98 & Yan et al. 1998 & C {\sc iv} \\
ZB09 & Zatsarinny \& Bartschat 2009 & Kr {\sc i} \\
ZBG09 & Zatsarinny et al. 2009 & Mg {\sc i} \\
ZBR00 & Zinner et al. 2000 & Ca {\sc i} \\
ZF02 & Zatsarinny \& Froese Fischer 2002 & C {\sc i} \\
\enddata

\end{deluxetable}

We encourage future studies of interstellar, circumgalactic, and intergalactic 
absorption lines to use the revised oscillator strengths compiled here. This is 
especially important for the $\sim 22\%$ of the lines where the improvements of 
$> 0.1$ dex are larger than the typically quoted measurement uncertainties in 
metal column densities (usually $< 0.05$ dex for non-saturated lines observed 
with state-of-the-art high-resolution spectrographs used for such studies, e.g., 
Keck High-resolution Spectrograph, VLT UV Echelle Spectrograph, {\it Magellan} Inamori 
Kyocera Echelle spectrograph). It would also be useful to confirm the oscillator 
strengths of the 43 transitions of \ion{C}{1}, \ion{Si}{1}, \ion{P}{2}, 
\ion{S}{1}, \ion{Cl}{1}, \ion{Ti}{1}, \ion{Ti}{2}, \ion{Mn}{1}, \ion{Mn}{2}, 
\ion{Fe}{2}, \ion{Ni}{2}, \ion{Kr}{1}, and \ion{Pb}{2} for which the differences 
in the oscillator strengths are $\gtrsim 0.2$ dex. Such improvements in atomic 
data are crucial to obtain accurate element abundances in distant galaxies, 
which are needed to quantitatively test models of cosmic chemical evolution. 
Indeed, improved atomic data will be invaluable for accurately interpreting the 
large samples of spectra of high-redshift quasars, gamma-ray bursts, and 
star-forming galaxies that will be enabled by future extremely large telescopes.

\newpage


\renewcommand{\baselinestretch}{0.95}
\noindent
\begin{deluxetable}{lll}
\tablecolumns{3} 
\tablewidth{0pt}
\tablecaption{
Statistical Analysis of Change in Oscillator Strength} 
\tablehead{ 
\colhead{$\Delta log f$ (dex)} &
\colhead{No. of lines} & 
\colhead{\% lines}
\label{tab:dex}
}
\startdata 
$\Delta log f$    $\le 0.05 $          & 198  & 58.2 \\
$0.05 <$  $\Delta log f$    $\le 0.1$  & 55   & 16.2 \\
$0.1  <$  $\Delta log f$    $\le 0.2$  & 35   & 10.3 \\
$0.2  <$  $\Delta log f$    $\le 0.3$  & 15   & 4.4  \\
$0.3  <$  $\Delta log f$    $\le 0.4$  & 6    & 1.8  \\
$\Delta log f$    $> 0.4    $          & 19   & 5.6  \\
no change                   & 12   & 3.5 \\
\enddata

\tablecomments {
This table compares the former and current oscillator strength values for 340 lines. 
The remaining 236 lines were either newly tabulated by the authors 
or no improved update was found for the former value.
}

\end{deluxetable}

\acknowledgments
We thank an anonymous referee for constructive suggestions on an earlier version of this manuscript. This work was supported by the collaborative National Science Foundation grants 
AST/1109061 to Univ. of Kentucky and AST/1108830 to Univ. of South Carolina. 
F.H.C. acknowledges partial support from a NASA/SC Space Grant graduate fellowship.
V.P.K. also acknowledges partial support from STScI ($HST$-GO-12536, $HST$-GO-13801, 
$HST$-GO-13733, $HST$-GO-14137) and NASA (NNX14AG74G). 
G.J.F. acknowledges support by NSF (1412155), NASA (ATP13-0153), and STScI 
($HST$-AR-13245, GO-12560, $HST$-GO-12309, GO-13310.002-A, $HST$-AR-13914, 
$HST$-AR-14286.001 and $HST$-AR-14556).

\end{document}